\documentclass[prl,aps,epsfig,longbibliography,superscriptaddress,twocolumn]{revtex4-1}
\usepackage{graphicx}
\usepackage[unicode=true,colorlinks=true]{hyperref}
\hypersetup{linkcolor=blue,citecolor=blue,urlcolor=blue}
\usepackage{dcolumn}
\usepackage{bm}
\usepackage{ulem}
\usepackage{color, colortbl}

\usepackage{color,url,nicefrac,multirow,amsmath,amssymb,amsfonts,tabularx,bm}
\usepackage[titletoc,title]{appendix}
\newcommand*{\rom}[1]{\expandafter\@slowromancap\romannumeral #1@}

\renewcommand{\vec}[1]{{\textbf{\textit{#1}}}}

\begin{document}
\title{An Exactly Solvable Model for Strongly Interacting Electrons in a Magnetic Field}
\begin{abstract}

States of strongly interacting particles are of fundamental interest in physics, and can produce exotic emergent phenomena and topological structures. We consider here two-dimensional electrons in a magnetic field, and, departing from the standard practice of restricting to the lowest LL, introduce a model short-range  interaction that is infinitely strong compared to the cyclotron energy. We demonstrate that this model lends itself to an exact solution for the ground as well as excited states at arbitrary filling factors $\nu<1/2p$ and produces a fractional quantum Hall effect at fractions of the form $\nu = n/(2pn + 1)$, where $n$ and $p$ are integers. The fractional quantum Hall states of our model share many topological properties with the corresponding Coulomb ground states in the lowest Landau level, such as the edge physics and the fractional charge of the excitations. 
\end{abstract}
\author{Abhishek Anand}
\affiliation{Indian Institute of Science Education and Research, Pune 411008, India}
\author{J K Jain}
\affiliation{The Pennsylvania State University, 104 Davey Laboratory, University Park, Pennsylvania 16802, USA}
\author{G J Sreejith}
\affiliation{Indian Institute of Science Education and Research, Pune 411008, India}

\maketitle

Spectacular phenomena arising from strong correlations are of fundamental interest in all areas of physics, and can sometimes produce topological phases of matter with exotic emergent properties. Exactly solvable models of strongly correlated systems are therefore of particular significance. We consider here the phenomenon of fractional quantum Hall effect (FQHE)~\cite{Tsui82}, namely quantization of Hall resistance at $R_{H}=h/fe^2$ where $f$ is a fraction, that occurs when two-dimensional (2D) electrons are subject to a strong magnetic field.  Close to a hundred fractions have been observed to date in various 2D electron systems in semiconductor quantum wells and graphene.  Several themes in contemporary condensed matter physics, such as fractional charge~\cite{Laughlin83} and fractional statistics~\cite{Halperin84}, composite fermions~\cite{Jain89}, non-Fermi liquids~\cite{Halperin93}, topological superconductivity with Majorana particles~\cite{Read00,Stern10}, and proposals for topological quantum computation~\cite{Nayak08}, have originated in the context of FQHE. This Letter presents a model of strongly correlated electrons in a magnetic field that is exactly solvable for all eigenstates at arbitrary filling factors $\nu<1/2p$, 
and produces FQHE at fractions of the form $f= n/(2pn + 1)$, where $n$ and $p$ are integers.

Theoretical investigations of FQHE begin with the Hamiltonian of 2D electrons in a magnetic field:
\begin{equation}
\hat{H}=\sum_{j=1}^N{\hat{\boldsymbol\pi}_j^2\over 2m_b} + \sum_{j<k=1}^N \hat{V}(|\vec{r}_j -\vec{r}_k| ),
\label{eq:Hdisk1}
\end{equation}
where $\hat{\boldsymbol\pi}_j=\hat{\bf p}_j+{e\over c}\vec{A}(\hat {\bf r}_j)$ is the kinetic momentum, $\vec{A}(\vec{r})=(B/2)(-y, x,0)$ is the vector potential producing a magnetic field $B$ in the $+{z}$ direction, $m_b$ is the electron band mass, $N$ is the number of electrons, and $\hat{V}$ is the interaction between the electrons. The eigenfunctions of the single particle Hamiltonian ${ \boldsymbol\pi}^2/2m_b$ are given by
\begin{equation}
\phi_{n,m}=e^{-\frac{r^2}{4}}z^m L_n^m\left(\frac{r^2}{2}\right),\;\; E_{n,m}=\left(n+\frac{1}{2}\right)\hbar\omega_c
\label{eq:singleparticle}
\end{equation}
where $n=0,1,2, \dots$ are called Landau levels (LLs), $m=-n, -n+1, \dots$ is  the angular momentum ($z$-component), $\hbar\omega_c=\hbar eB/m_bc$ is the cyclotron energy;  $L_n^m$ is the associated Laguerre polynomial; and $z=x-\imath y$. All lengths are in units of magnetic length $\ell = \sqrt{\hbar c/eB}$. An important parameter is the filling factor, defined as $\nu=N/N_\phi$, where $N$ is the number of particles and $N_\phi=BA/\phi_0$ is the number of flux quanta ($\phi_0=hc/e$) through the sample area $A$. The quantized plateaus with $R_{H}=h/fe^2$ occur in a range of filling factors around $\nu=f$.

The traditional theoretical practice since the early works~\cite{Laughlin83,Haldane83} has been to consider the limit where the interaction energy is weak compared to the cyclotron energy, i.e., $\kappa\equiv V_C/\hbar\omega_c\rightarrow 0$, where $V_C=e^2/\epsilon \ell$ is the Coulomb interaction scale, $\epsilon$ being the dielectric constant of the host material. The reason the system becomes strongly correlated is that, in this limit, electrons are confined to the lowest LL (LLL), leaving the interaction as the only energy scale in the problem. The problem with Coulomb interaction is not exactly solvable. Haldane introduced~\cite{Haldane83} a model interaction that obtains the Laughlin wave function at $\nu=1/3$ as the exact ground state (GS). Other model interactions have been designed for several known trial wave functions~\cite{Trugman85,Jain90b,Jain90,Rezayi91,Greiter91,Simon07b,Wu17,Bandyopadhyay18,Bandyopadhyay20,Simon20}. These models, however, are not fully solvable, in that each model produces incompressibility only at one fraction, and only the zero energy solutions are known. 

In this Letter, we introduce a model interaction that can be solved exactly for all eigenstates at arbitrary filling factors $\nu<1/2p$ and 
yields FQHE at all fractions of the form $n/(2pn+1)$. A fundamental departure from the previous approaches is that we consider the opposite limit where the interaction is infinitely strong compared to the cyclotron energy, leaving the cyclotron energy as the only energy scale in the problem. The role of the interaction is to impose a nontrivial constraint on the allowed (finite energy) space of many-particle wave functions, which is responsible for FQHE within our model.

\begin{figure}
\includegraphics[width=0.9\columnwidth]{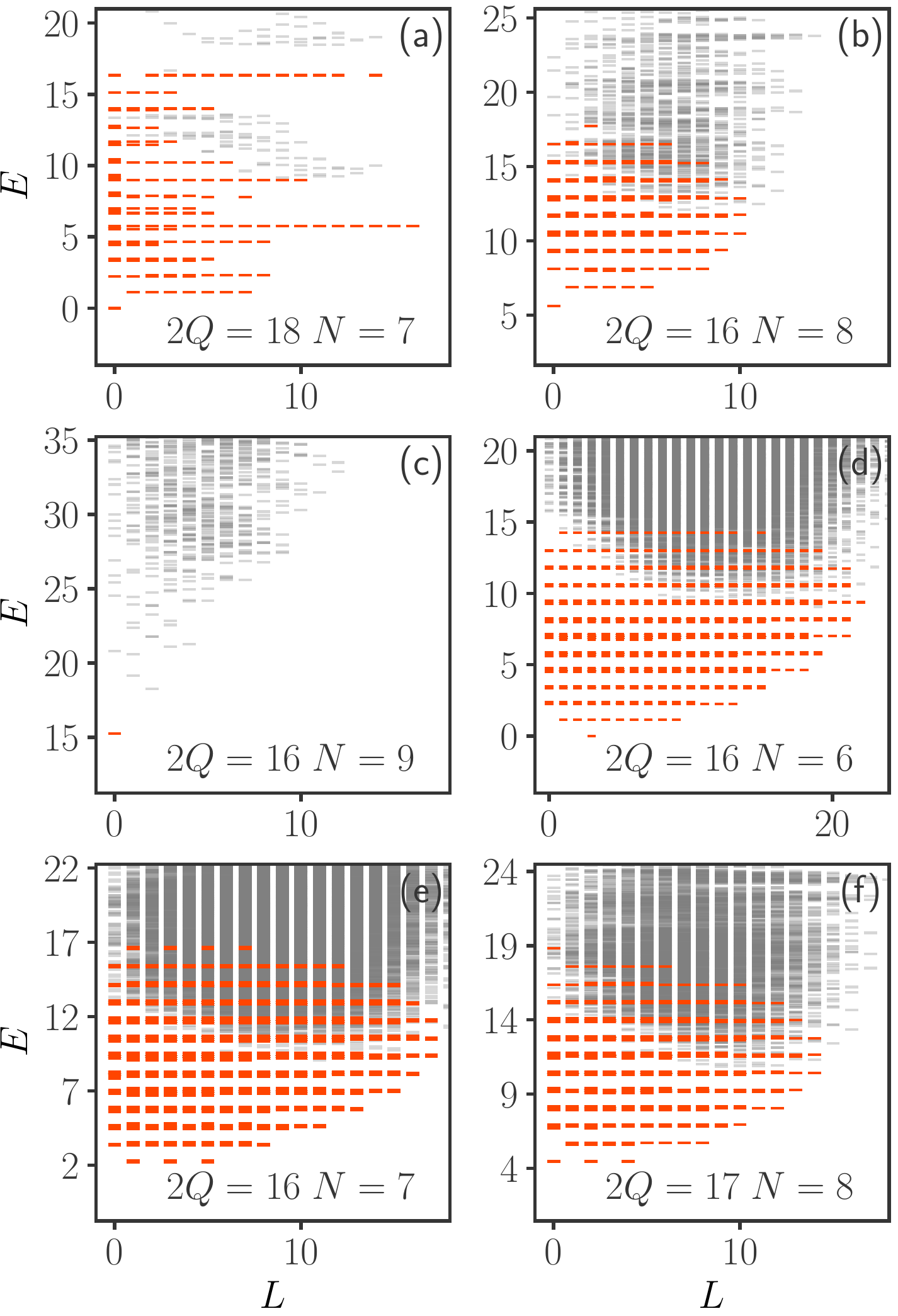}
\caption{
Spectra of the model Hamiltonian for representative systems in the spherical geometry, with $N$ particles in $2Q$ flux quanta. All nonzero pseudopotentials [Eq.(\ref{Hsphere})] are set to $V_{L}^{n,n'}=20\hbar\omega_c$, and the Hilbert space is restricted to the three lowest LLs. The energy $E$ is in units of $\hbar \omega_c$, measured relative to the zero point energy $N\hbar\omega_c/2$. $L$ is the total orbital angular momentum. Orange markers show the $\mathcal{V}_\infty$ states with zero interaction energy. (Each orange dash may represent many degenerate eigenstates.) Black dashes show eigenstates with nonzero interaction energies; these will be pushed to infinity in the limit $V_{L}^{n,n'}/\hbar\omega_c\rightarrow \infty$.
(a)--(c) Spectra for 
systems at $\nu=1/3$, $2/5$, and $3/7$; for $\nu=3/7$ the spectrum has no orange-colored excited states, as we have kept only the three lowest LLs in our study. (d)--(f) Spectra for $N$ and $2Q$ for which the GS is not incompressible.
Panel (a) shows only a subset of the higher energy states due to difficulty in diagonalization of Hamiltonians with large degeneracies. The slight broadening of the orange bands is a finite size effect, resulting from the fact that for finite systems in the spherical geometry, the inter-LL gap depends slightly on the LL index.
\label{fig:spectr}}
\end{figure}

\textit{Model Hamiltonian.---} 
Our model interaction is inspired by composite fermion (CF) physics~\cite{Jain89}. A CF is the bound state of an electron and $2p$ vortices. The FQHE arises due to the integer quantum Hall effect (IQHE) of CFs. The electronic wave function is obtained by multiplying the IQH state by the Jastrow factor $\prod_{j<k}(z_j-z_k)^{2p}$, which converts electrons into CFs by vortex (or flux) attachment, and also increases the {\it relative} angular momentum of each electron pair, denoted $M$, by $2p$ units.  

We proceed as follows. (We assume fully spin polarized electrons; generalization to spinful electrons or bosons is straightforward.) For two electrons in the $n$th and $n'$th LLs, the smallest value of $M$ is $-n-n'+\delta_{n,n'}$. We next increase $M$ of each pair by $2p$, {\it but in a manner that preserves their LL indices}. In the resulting state the smallest $M$ for a pair with one electron in the $n$th and the other in $n'$th LLs is $-n-n'+\delta_{n,n'}+2p$. In other words, pairs with $-n-n'+\delta_{n,n'} \leq M < -n-n'+\delta_{n,n'}+2p$ are absent. This motivates us to construct an interaction that imposes an energy penalty on these pairs:
\begin{equation}
\hat{V}= \sum_{n\leq n'=0}^{\infty}\;\sum_{M=-n-n'+\delta_{n,n'}}^{-n-n'+\delta_{n,n'}+2p-1} V^{n,n'}_M |n,n';M\rangle \langle n,n';M|.
\label{eq:Hdisk2}
\end{equation}
Here $|n,n';M\rangle$ represents the state of a pair of electrons, from the $n$th and $n'$th LLs, with relative angular momentum $M$, and $V^{n,n'}_M=  \langle n,n';M| \hat{V}  |n,n';M\rangle  $ is the interaction energy of this pair, a generalization of the Haldane pseudopotentials including intra- as well as inter-LL pairs. (Note that pairs with certain $M$'s may not be allowed due to exchange symmetry; the corresponding terms are automatically absent in the Hamiltonian. We have suppressed, for ease of notation, the center-of-mass degree of freedom, which does not affect the pseudopotentials.)  The above interaction conserves the number of electrons in each LL (denoted $N_n$ for the $n$th LL) and therefore kinetic energy. The total angular momentum $m_{\rm total} =\sum_j m_j$, and $(N_0,N_1,N_2,\dots)$ label each eigenstate. To describe the low-energy physics of a given fraction at $\nu=\nu^*/(2p\nu^*+1)$, terms containing only a finite number of LLs need to be kept in the above summation; this restricted interaction is thus local. 

\textit{$\mathcal{V}_\infty$ wave functions.---} 
We will now consider the strong interaction limit ${V^{n,n'}_m/\hbar \omega_c}\rightarrow \infty$, so all eigenstates with nonzero interaction energy are projected out. 
The collection of wave functions that (i) have zero interaction energy and (ii) are eigenstates of the kinetic energy will be referred to as the ${\cal V}_\infty$ wave functions. The above physics motivates the following construction of ${\cal V}_\infty$ wave functions.
We denote by $\Phi^\alpha$ the distinct kinetic energy eigenstates, labeled by $\alpha$, of noninteracting fermions; these are simple Slater determinants. To construct the ${\cal V}_\infty$ states, we must increase the relative angular momentum of each pair by $2p$ units in a LL-conserving manner. The standard composite fermionization through multiplication by $\prod_{j<k}(z_j-z_k)^{2p}$ does not conserve the LL index. Instead consider: 
\begin{equation}
\Psi^\alpha=\prod_{j<k}(\hat{Z}_j-\hat{Z}_k)^{2p}\times \Phi^\alpha
\label{eq:Psinu*}
\end{equation}
where $\hat{Z}=\hat{z}-\imath (\hat{\pi}_x-\imath \hat{\pi}_y)/\hbar$ is the guiding center coordinate. 
That $\Psi^\alpha$ is a ${\cal V}_\infty$ state follows from two facts: (i)  The guiding center coordinate operator $\hat{Z}$ is similar to the position operator but without any 
matrix elements that scatter between LLs. As a result, $\hat{Z}$ conserves the LL indices of the particles. $\Psi^\alpha$ thus has the same kinetic energy as $\Phi^\alpha$. (ii)  The angular momentum operator for a single particle [Eq.(\ref{eq:singleparticle})] can be written as $\hat{m}=\hat{Z}\hat{Z}^\dagger/2 - \hat{n}$, $\hat{n}$ being the LL index. The guiding center coordinate acts as a raising operator for $\hat{m}$ due to the commutation relation $[\hat{m},\hat{Z}]=\hat{Z}$. Thus $\hat{Z}$ raises the single-particle angular momentum. In the symmetric gauge, the guiding center coordinate has a real space representation of $\hat{Z}\equiv z/2-2\partial_{\bar{z}}$. For a pair of particles, $\hat{Z}_1-\hat{Z}_2$ does not change the center-of-mass angular momenta, because it commutes with the center of mass $(\hat{z}_1+\hat{z}_2)/2$ and its angular momentum.
The Jastrow operator $\prod_{j<k}(\hat{Z}_j-\hat{Z}_k)^{2p}$ thus increases the relative angular momentum of every pair of particles by $2p$.  Because $\Psi^\alpha_{\nu}$ does not contain pairs with relative angular momenta for which $\hat{V}$ imposes a penalty, $\Psi^\alpha$ has zero interaction energy.

The action of the Jastrow operator increases the largest occupied single-particle angular momentum by $2p(N-1)$ such that the state $\Phi^\alpha$ at $\nu^*$ produces a state $\Psi^\alpha$ at $\nu=\nu^*/(2p\nu^*+1)$ \cite{Jain89,Jain07}.  
We next make the conjecture that the states in Eq.(\ref{eq:Psinu*}) provide a complete basis for the ${\cal V}_\infty$ space.  This conjecture 
implies that the excitation spectrum of our model Hamiltonian at any filling $\nu=\nu^*/(2p\nu^*+1)$ is identical to that of noninteracting fermions at filling $\nu^*$. Extensive diagonalization studies, discussed below, provide a compelling and nontrivial 
confirmation of the completeness of $\Psi^\alpha_{\nu}$.  As a corollary, our model produces FQHE at $\nu= n/(2pn+1)$, in analogy with the IQHE at $\nu^*=n$.  

\textit{Exact spectrum on sphere.---}For diagonalization studies, we find it convenient to employ Haldane's spherical geometry~\cite{Haldane83}, where $N$ electrons move on a sphere with a total radial magnetic flux of $2Q\phi_0$, where $2Q$ is an integer. The single-particle orbitals in the $n$th LL have angular momentum $l=|Q|+n$. The eigenstates are labeled by the total angular momentum $L$. The relative pair angular momentum $M$ of the disk geometry corresponds to pair angular momentum $L=2Q-M$ on the sphere. The interaction [Eq.(\ref{eq:Hdisk2})] therefore translates to
\begin{equation}
\hat{V}= \sum_{n\leq n'=0}^{\infty}\;\sum_{L=2Q+n+n'-\delta_{n,n'}-2p+1}^{2Q+n+n'-\delta_{n,n'}} V^{n,n'}_L |n,n';L\rangle \langle n,n':L|.
\label{Hsphere}
\end{equation}
According to the standard CF theory the interacting system $(N, 2Q)$ of electrons relates to a noninteracting system $(N, 2Q^*)$ of CFs with $Q^*=Q - p(N-1)$. We specialize to $2p=2$ below.

\begin{figure}[h!]
\includegraphics[width=0.9\columnwidth]{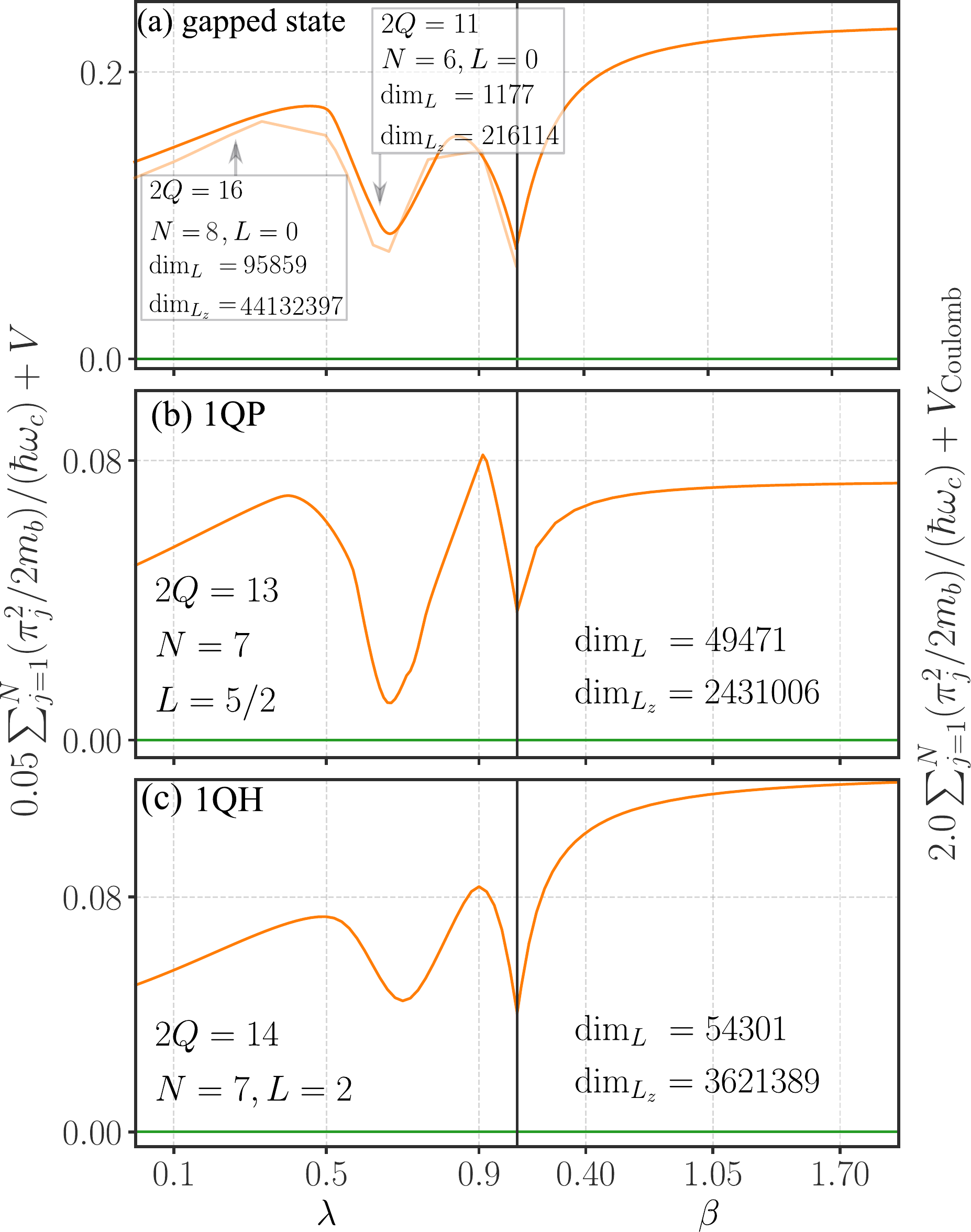}
\caption{
Adiabatic connection between the model and Coulomb Hamiltonian at $\nu=2/5$. The Hamiltonian is defined by $H=\frac{\beta}{\hbar \omega_c}\sum_{j=1}^N{\hat{\boldsymbol\pi}_j^2\over 2m_b}+(1-\lambda) V+\lambda V_{\rm Coulomb}$, where all energies are in units of $e^2/\epsilon \ell$, $V$ is our model interaction in which all nonzero pseudopotentials are set to unity, and $V_{\rm Coulomb}$ is the Coulomb interaction.
Panel (a) shows the lowest two energies (green, orange) in the $L=0$ sector in a system with an incompressible GS as the Hamiltonian is varied. On the left side, the interaction Hamiltonian is varied (parametrized by $\lambda$) from the model interaction to the Coulomb interaction, keeping $\beta=0.05$. On the right side, the cyclotron energy (parametrized by $\beta$) is sent to a large value ($\beta=2$), with the interaction fixed at its Coulomb value. All energies are measured relative to the GS energy. The darker (lighter) shade line shows the data for $N=6$ ($8$) particles. For $N=8$, evolution in only the initial half was calculated. Panels (b) and (c) correspond to the evolution of a single QP and a single QH state of $2/5$ that occur at angular momenta $5/2$ and $2$. In all cases, the gap is seen to remain finite throughout. In row (a), at small $\lambda$, the simplest excitation in the $L=0$ corresponds to two particle-hole pairs whose energy above ground state is given by $2 \beta (1+1/Q)$; the deviation from $2\beta$ arises because, in the spherical geometry, the LL separation has a finite size deviation from $\hbar\omega_c$. The dimensions of the Hilbert space in the relevant $L$ and $L_z$ sectors are shown on the figures, denoted dim$_L$ and dim$_{L_z}$, respectively.
\label{fig:adiabatic25}
}
\end{figure}
We have performed exact diagonalization of this Hamiltonian for many systems $(N,2Q)=(5,12)$, $(5,11)$, $(6,11)$, $(6,12)$, $(6,15)$, $(6,16)$, $(7,16)$, $(8,16)$, $(9,16)$, $(8,17)$, and $(7,18)$ in the Hilbert space restricted to the three lowest LLs. See  the Supplemental Material ~\cite{SM-ExactlySolvable20} for details. In many cases the full Hilbert space is prohibitively large, but the calculation is possible because diagonalization can be performed in each $(N_0,N_1,N_2)$ sector separately. The resulting spectra are shown in Fig.~\ref{fig:spectr} and the Supplemental Material. In our calculations we have set all nonzero pseudopotentials to unity and 
$\hbar\omega_c=0.05$. The states with zero interaction energy (in the $\cal{V}_\infty$ sector) are highlighted in orange. 
The high-energy parts of the spectra in Fig.~\ref{fig:spectr} also show states outside the ${\cal V}_\infty$ space (black dashes), which will be pushed to infinity in the limit ${V^{n,n'}_m/\hbar \omega_c}\rightarrow \infty$.

The states in the $\cal{V}_\infty$ sector display IQHE-like structure of bands of states separated by $\sim \hbar\omega_c$ ($\sim$ is needed because the LL spacing in the spherical geometry is not exactly $\hbar\omega_c$ for finite systems). In fact, for each case, including those shown in Fig.~\ref{fig:spectr}, the spectrum is identical to that for $N$ noninteracting electrons at flux $Q^*=Q-p(N-1)$ (as shown explicitly in the Supplemental Material).  The restriction to three LLs is purely for feasibility of numerics and is not a limitation for the ansatz solutions.
We have tested close to 200 different $(N_0,N_1,N_2, 2Q)$ systems~\cite{SM-ExactlySolvable20}, and in each case the dimension of the ${\cal V}_\infty$ space in the exact spectrum matches that of noninteracting fermions at $(N_0,N_1,N_2, 2Q^*)$. 
We thus have exact wave functions and energies not just for the incompressible GSs but also for its quasiholes (QHs), quasiparticles (QPs), neutral excitations, and, in fact, all ${\cal V}_\infty$ eigenstates. [QHs (QPs) of $\Psi^\alpha$ in Eq.(\ref{eq:Psinu*}) correspond to holes (particles) in $\Phi^\alpha$  in an otherwise full (empty) LL.]  The exact mapping to noninteracting fermions at filling $\nu^*$ is lost if we set some of the $V^{n,n'}_L$ in Eq.(\ref{Hsphere}) to zero or include additional pseudopotentials; in particular, the inter-LL pseudopotentials are necessary.

\textit{Topological properties---}
While $\kappa$ was small in the early experiments~\cite{Tsui82}, FQHE is routinely seen for $\kappa \sim 5-7$ in \textit{p}-doped GaAs~\cite{Santos92} and ZnO quantum wells~\cite{Maryenko18} and is seen to survive even up to $\kappa\sim 40$ in high-quality AlAs quantum wells~\cite{Shayegan20b}. The limit of strong interactions is thus not entirely unphysical for FQHE. While our model is not intended to produce a quantitative theory of experiments for large $\kappa$, we now show that it shares topological features with the FQHE of Coulomb electrons in the LLL. 

Our theory reproduces the prominent qualitative features of the phenomenology, namely gapped states at $\nu=n/(2pn+ 1)$ and, by extension to the $n\rightarrow \infty$ limit, the compressible Fermi sealike states~\cite{Halperin93} at $\nu=1/2p$. {Many topological properties follow from the counting of lowest energy excitations, which our model produces exactly.} The fractional charge of magnitude $e^*=e/(2pn+1)$ for the QPs and QHs follows from the presence of gap at $\nu=n/(2pn+ 1)$ but can also be deduced from the mapping into $2Q^*=2Q-2p(N-1)$\cite{SM-ExactlySolvable20,Jain07}. We have confirmed fractional charge by explicit evaluation in finite-size systems (Supplemental Material \cite{SM-ExactlySolvable20}). The one-to-one correspondence with IQHE implies that the excitations obey Abelian braid statistics, because the wave function $\Phi^\alpha$, and thus also $\Psi^\alpha$, is uniquely determined by specifying the positions of the QPs or QHs. The edge physics of the FQHE state $\nu=n/(2pn+1)$ is analogous to that of the IQHE state at $\nu^*=n$, described by $n$ chiral edge modes.  Real space or momentum space entanglement spectra should also reflect these edge modes \cite{LiHaldane}, but the currently accessible system sizes are insufficient to reveal the characteristic features (except for $\nu=1/3$)\cite{EntSpecOf25Gaffnian2009,EntSpecLargeSystem25th2013}. Another topological quantity is the shift $S$, or the orbital spin~\cite{Wen92}, defined by the relation $2Q=\nu^{-1}N-S$ for the flux where FQHE states occur in the spherical geometry. Our model produces the same shift, $S=n+2p$, as the standard CF theory. 

We next ask whether a path exists in the space of Hamiltonians that takes the gapped states of the model to those of the LLL Coulomb Hamiltonian without gap closing. This parameter space is large, consisting of the cyclotron energy and many pseudopotentials, leaving us with many different paths to consider. If we simply increase the cyclotron energy while retaining the model interaction, the orange states in Fig.~\ref{fig:spectr} will float up, resulting in many level crossings. (The only exceptions are the GS and QH states at $\nu=1/3$, which are already in the LLL, and also ground states of the $V_1$ pseudopotential;  these remain ground states throughout.) However, a more general interaction will convert the level crossings into anticrossings, for any finite system, thus opening a gap everywhere. We ask if we can identify a path where the gap remains robust throughout. We have studied the evolution of the GS, the QP state and the QH state as the Hamiltonian is continuously deformed from our model to the Coulomb interaction with large cyclotron energy for $\nu=1/3$ and $\nu=2/5$.
Figure \ref{fig:adiabatic25} shows the evolution of the incompressible GS, the single QP state, and the single QH state at filling factor $\nu=2/5$ along the following path: we first vary parameters to go from our model interaction (all nonzero pseudopotentials set to unity) continuously to Coulomb (left side), and then increase $\hbar\omega_c$ (right side). The GS, single QP and single QH states of the model evolve continuously into the corresponding LLL Coulomb states. The adiabatic continuity for the GS, QH and QP of $\nu=1/3$ is shown in the Supplemental Material \cite{SM-ExactlySolvable20}. 
For the largest systems at $1/3$ and $2/5$ that we could diagonalize, the physics of the model and the LLL Coulomb Hamiltonian are adiabatically connected. Nonetheless, as with all numerical results, we cannot definitively assert that adiabatic continuity will survive in the thermodynamic limit. 

In summary, we report on an exactly solvable model for a strongly correlated system of electrons in a magnetic field. An appealing aspect of this model is that it exactly implements the physics of ``noninteracting CFs," in that the spectrum of strongly interacting electrons at $\nu=\nu^*/(2p\nu^*+1)$ matches that of noninteracting fermions at $\nu^*$. 
Although the interaction is presented employing the relative angular momentum quantum numbers, which requires rotational symmetry, the interaction is local and its physics is expected to be independent of the boundary conditions in the thermodynamic limit. Generalization  to the torus geometry\cite{ManyBodyTranslationsTorus,PeriodicLaughlin,Maria2013,Pu2017} is a natural question in this context and will be addressed in a future publication. The wave functions in this work can be  extended to accommodate an anisotropic electron mass\cite{GemoetricDescriptionHaldane,anisotropicQHS}; construction of a suitable anisotropic interaction for these states is an interesting open question.

\begin{acknowledgments}
G. J. S. acknowledges financial support from DST-SERB (India) Grant No. ECR/2018/001781 and a joint grant from IISER-Pune CNRS. 
J. K. J. acknowledges financial support from the U.S. Department of Energy, Office of Science, Basic Energy Sciences, under Award No. DE-SC0005042. 
Calculations were performed with PETSc and SLEPc libraries in the Param Brahma (NSM, IISER Pune) and Param Yuva II (CDAC) supercomputing facilities.
G. J. S. thanks Fabien Alet for useful discussions. 
\end{acknowledgments}


\clearpage
\pagebreak

\newcommand{\CalF}{{\mathcal{F}}}
\newcommand{\NSymPol}{N}
\newcommand{\NSDTotal}{N}
\newcommand{\NSDGroup}{\mathcal{N}}
\newcommand{\Nsphere}{{{N}}}
\newcommand{\Nb}{{{N_b}}}
\newcommand{\LAngMom}{{\mathcal{L}}}
\newcommand{\LLockZero}{{L}}

\renewcommand*\arraystretch{1.5}

\setcounter{figure}{0}
\setcounter{table}{0}
\setcounter{equation}{0}
\renewcommand\thefigure{S\arabic{figure}}
\renewcommand\thetable{S\arabic{table}}
\renewcommand\theequation{S\arabic{equation}}
\setcounter{secnumdepth}{4}

\begin{center}
\textbf{Supplemental Material}
\end{center}

In Sec.~\ref{SSecI} we introduce guiding center coordinates. The Haldane pseudopotentials are generalized to the problem including many LLs in Sec.~\ref{SSecII}. This is followed by a discussion of the matrix elements for the Coulomb interaction as well as our model interaction in Section \ref{SSecIII}. In Sec.~\ref{SSecIV} we present additional spectra from exact diagonalization of our model Hamiltonian. Sec.~\ref{SSecV} gives a large number of examples showing that the counting of states in exact diagonalization matches precisely with our conjecture that maps it into a system of noninteracting fermions. Sec.~\ref{SSecVI} connects our model with the LLL Coulomb model for 6 particles at $\nu=1/3$ by demonstrating adiabatic continuity between the two. (Similar relation is demonstrated for $\nu=2/5$ in the main text.) 
In section \ref{SSecQP}, we present the results for the charge distribution around simple excitations of $1/3$ and $2/5$ states.
We conclude with a discussion of the Trugman-Kivelson model interaction, which obtains a related wave function as the exact ground state at $\nu=2/5$.

\section{Landau quantization and guiding center coordinates}\label{SSecI}

The Hamiltonian of a particle of mass $m$ and change $-e$, in a uniform magnetic field in the $z$ direction is given by
\begin{equation}
\hat{H} = \frac{{\hat{\boldsymbol \pi}}^2}{2m_b},\;\text{ where}\; \hat{\boldsymbol \pi} = \hat{\bf p} + \frac{e}{c}{\bf A}(\hat{\bf r}),
\label{SHkin}
\end{equation}
The kinetic momentum $\hat{\boldsymbol \pi}$ satisfies the commutation relations $[\hat{x}_j,\hat{\pi}_i]=\imath \hbar \delta_{ij}$ similar to the canonical momentum. The components of $\hat{\boldsymbol \pi}$ however do not commute; we have $[\hat{\pi}_x,\hat{\pi}_y]=-\imath \frac{\hbar^2}{\ell^2}$ where $\ell$ is the magnetic length $\sqrt{\hbar c/eB}$. The Hamiltonian in Eq.~\ref{SHkin} is a thus quadratic function of canonically conjugate variables, reminiscent of the quantum harmonic oscillator. Following arguments analogous to those in the quantum harmonic oscillators, the Hamiltonian can be expressed as $\hbar\omega_c (\hat{a}^\dagger \hat{a}+1/2)$ where the energy ladder operator defined as
\begin{equation}
\hat{a}^\dagger = \frac{\ell}{\sqrt{2}\hbar}\left(\hat{\pi}_x + \imath \hat{\pi}_y\right)
\end{equation}
satisfies a commutation relation $[\hat{a},\hat{a}^\dagger]=1$. The operator raises the energy of a state it acts on by $\hbar \omega_c$ on account of its commutation relation with the Hamiltonian: $[H,\hat{a}^\dagger] = \hbar \omega \hat{a}^\dagger$. The cyclotron frequency $\omega_c$ is given by ${eB}/{m_bc}$.

Action of the energy ladder operator covers only part of the degrees of freedom. This can be seen upon exploring the dynamics of the position operator $\hat{z}=\hat{x}-\imath \hat{y}$.
\begin{multline}
\partial_t \hat{z} = \frac{1}{\imath \hbar} [\hat{z},\hat{H}] = \frac{1}{m}(\hat{\pi}_x-\imath \hat{\pi}_y) = \frac{\sqrt{2}\hbar}{m\ell}\hat{a}
\end{multline}
The time evolution of the position resembles the time evolution of $a$:
\begin{equation}
\partial_t \hat{a} = \frac{1}{\imath \hbar} [\hat{a},H] = -\imath \omega_c \hat{a}
\end{equation}
These suggest that a suitable linear combination of $\hat{z}$ and $a$, namely 
\begin{equation}
\hat{Z} = \hat{z} -  \imath \sqrt{2} \ell \hat{a},
\end{equation} 
is time independent, i.e. commutes with the Hamiltonian. This constant of motion is called the guiding center coordinate. Action of this operator changes a state without altering its energy, indicating that it affects degrees of freedom different from those captured by the the energy ladder operators. It is easily verified that $\hat{Z}$ and $\hat{a}$ together satisfy the following commutation relations
\begin{equation}
[\hat{a},\hat{a}^\dagger]=1;\; [\hat{a},\hat{Z}]=[\hat{a}^\dagger,\hat{Z}]=0;\; [\hat{Z}^\dagger,\hat{Z}] = 2\ell^2
\end{equation}
The space in which this algebra acts is spanned by the energy states of the form
\begin{equation}
\left | n,s-n \right \rangle=\frac{(\hat{a}^\dagger)^n}{\sqrt{n!}} \frac{(\hat{Z}/\sqrt{2}\ell)^s}{\sqrt{s!}} \left | 0,0 \right \rangle;
n,s=0,1,\dots\infty
\end{equation}
where $|0,0\rangle$ is annihilated by both $\hat{a}$ and $\hat{Z}^\dagger$. This state has an energy $(n+1/2)\hbar \omega_c$ and angular momentum $m=s-n$. 
The LL index $n=0,1,2\dots $ and the angular momentum $m=-n,-n+1,\dots \infty$  
are eigenvalues of the LL index operator $\hat{n}=\hat{a}^\dagger \hat{a} $ and the angular momentum operator $\hat{m}=\hat{Z}\hat{Z}^\dagger/2\ell^2 -\hat{n}$. 
The guiding center coordinates $\hat{Z}$ and $\hat{Z}^\dagger$ act as raising and lowering operators for the angular momentum.

The discussion so far is gauge independent. Guiding center coordinate $\hat{Z}$ can be written in real space representation once a choice for the gauge $\bf{A}$ has been made. In the symmetric gauge ${\bf A}\equiv B(-y,x)/2$, the guiding center coordinate is 
\begin{multline}
\hat{Z} = \hat{z} -  \imath \sqrt{2}{\ell}\hat{a} =  \hat{z}  - \imath \frac{\ell^2}{\hbar} (\hat{\pi}_x - \imath \hat{\pi}_y) \\
\equiv \frac{x-\imath y}{2} - {\ell^2} (\partial_x - \imath \partial_y) = \frac{z}{2} - 2\ell^2 \partial_{\bar{z}}
\end{multline}

\section{Generalized Haldane pseudopotentials}
\label{SSecII}

Let us denote by $|n_1 n_2; \tilde{M}  M\rangle$ the state of two electrons that occupy the $n_1$th and $n_2$th LLs and have a relative angular momentum $M$ and center of mass angular momentum $\tilde{M}$. It is evident that the state labeled by these quantum numbers, if allowed by symmetry, is unique. It can be generated by exact diagonalization of a rotationally invariant interaction in the appropriate Hilbert space. In fact, given the wave function for the minimum relative angular momentum pair for a given $n_1$ and $n_2$, the states may be generated by repeated application of $(\hat{Z}_1\pm \hat{Z}_2)$. More specifically, we write 
\begin{multline}
|n_1 n_2; \tilde{M} M \rangle \equiv (\hat{Z}_1+\hat{Z}_2)^{\tilde{M}}  (\hat{Z}_1-\hat{Z}_2)^{{M}} \times \\  | (n_,-n_1)(n_2,-n_2+\delta_{n_1n_2})\rangle
\end{multline}
where 
$| (n_1,-n_1)(n_2,-n_2+\delta_{n_1n_2})\rangle=| n_1, -n_1\rangle  | n_2, -n_2+\delta_{n_1 n_2} \rangle $ 
is the two-particle state with the minimum relative angular momentum and zero center of mass momentum.

An interaction is characterized by the generalized Haldane's pseudopotentials 
\begin{equation}
V_M^{n_1 n_2; n_1' n_2'}=\langle  n_1' n_2'; \tilde{M} M|  \hat{V}(|\vec{r}_1-\vec{r}_2|)  |n_1 n_2; \tilde{M} M \rangle
\end{equation}
The pseudopotential does not depend on $\tilde{M}$, but, in general, has matrix elements involving LL transitions.  Our model interaction preserves LLs, and thus keeps only the diagonal pseudopotentials $V_M^{n_1 n_2; n_1 n_2}\equiv V_M^{n_1 n_2}$.

\section{Interaction Hamiltonians in the spherical geometry}
\label{SSecIII}

A general two-particle interaction between particles on a sphere can be written as follows
\begin{equation}
\sum_{l_1,l_2=|Q|}^\infty \sum_{m_1=-l_1}^{l_1} \sum_{m_2=-l_2}^{l_2} c_{l_1m_1}^\dagger c_{l_2m_2}^\dagger   V_{12;1'2'} c_{l'_2m'_2} c_{l'_1m'_1}
\end{equation}
where $V_{12;1'2'}$ is a shorthand for the matrix element of the interactions between two-particle states ie 
\begin{equation}
V_{12;1'2'}\equiv \langle (l_1,m_1),(l_2,m_2)  | \hat{V}  | (l'_1,m'_1),(l'_2,m'_2)\rangle\label{eq:defMatrixEl}
\end{equation}
where $| (l_1,m_1),(l_2,m_2)\rangle$ is a two-particle state with particles occupying $m_1$ and $m_2$ angular momentum states in the LLs $l_1$ and $l_2$ respectively.

Any two-particle interaction is fully characterized by its matrix elements. In the remainder of this section, we first describe the calculation of the matrix elements for the Coulomb interaction~\cite{2015APS..MAR.D5003W}, as well as for our model interaction, for electrons residing on the surface of a sphere. 

\subsection{Coulomb interaction}

The Coulomb interaction between particles on the surface of a sphere is $1/r$ where $r$ is the chord distance $|{\bf r}_a-{\bf r}_b|$ between the positions ${\bf r}_a \equiv (\theta_a,\phi_a)$ and ${\bf r}_2 \equiv (\theta_b,\phi_b)$.
The radius of the sphere is $R=\sqrt{Q}$ in units of the magnetic length $\ell=\sqrt{\hbar c/eB}$, $B=2Q\Phi_0/4\pi R^2$ is the field strength on the surface of the sphere, and $2Q$ is the (integer) number of flux quanta emanating from a monopole placed at the center of the sphere. Eigenstates of the single particle Hamiltonian $\hat{{\boldsymbol \pi}}^2/2m_b$ in this geometry are called the monopole spherical harmonics $Y_{Qlm}$.\cite{wu1977some,Wu:1976ge} The quantity $l=|Q|+n$ is the angular momentum of the orbitals in the $n$th LL ($n=0, 1, \dots$) and $m=-l,-l+1,\dots l$ labels the degenerate states in the $n$th LL .

The Coulomb potential can be expanded in terms of the Legendre polynomials as 
\begin{equation}
\frac{1}{r} = \frac{1}{\sqrt{Q}}\sum_{l=0}^\infty P_l(\cos \theta),\nonumber
\end{equation}
where $\theta$ is the angle subtended by ${\bf r}_a$ and ${\bf r}_b$ at the origin. The Legendre Polynomial can be expressed in terms of the monopole spherical harmonics (Eq.~14.30.9 of Ref.~\onlinecite{NIST:DLMF}) resulting in the expression 
\begin{equation}
\frac{1}{r} = \frac{1}{\sqrt{Q}}\sum_{l=0}^\infty \frac{4\pi}{2l+1} \sum_{m=-l}^l  Y^*_{0lm} (a) Y_{0lm} (b).\label{eq:multipoleexp}
\end{equation}
Here the arguments $a$ and $b$ of the monopole Harmonics are a short-hand notation to represent the coordinates $(\theta_a,\phi_a)$ and $(\theta_b,\phi_b)$.
The matrix element (Eq. \ref{eq:defMatrixEl}) of the Coulomb interaction between a pair of (antisymmetrized) two-particle fermionic states $\left|(l_1,m_1)(l_2,m_2)\right \rangle $ and $\left |(l'_1,m'_1)(l'_2,m'_2)\right \rangle$ is given by 
\begin{multline}
V_{12;1'2'}\equiv V_{(l_1m_1)(l_2m_2)}^{(l'_1m'_1)(l'_2m'_2)} - V_{(l_1m_1)(l_2m_2)}^{(l'_2m'_2)(l'_1m'_1)} +\\+ V_{(l_2m_2)(l_1m_1)}^{(l'_2m'_2)(l'_1m'_1)} - V_{(l_2m_2)(l_1m_1)}^{(l'_1m'_1)(l'_2m'_2)} \label{eq:antisymmet}
\end{multline}
where 
\begin{multline}
V_{(l_1m_1)(l_2m_2)}^{(l'_1m'_1)(l'_2m'_2)} = \\ \int  Y^*_{Ql_1m_1} (a) Y^*_{Ql_2m_2} (b) V(r) Y_{Ql'_1m'_1} (a) Y_{Ql'_2m'_2} (b)d\Omega_a d\Omega_b\label{eq:MainIntegral}.
\end{multline}
Theorems 1 and 3 from Ref \onlinecite{wu1977some} can be used to evaluate the integrals involved in terms of Wigner 3j symbols, yielding the following expression for $V_{(l_1m_1)(l_2m_2)}^{(l'_1m'_1)(l'_2m'_2)}$:
\begin{widetext}
\begin{multline}
\sum_{l=0}^\infty \nu_l \sum_{m=-l}^l \sqrt{(2l_1+1)(2l_2+1)(2l'_1+1)(2l'_2+1)} (-1)^{2Q+m_1+m_2+l_1+l_2+l'_1+l'_2} \\
\begin{pmatrix} l_1 & l & l'_1 \\ -m_1 & -m & m_1\end{pmatrix} \begin{pmatrix} l_2 & l & l'_2 \\ -m_2 & m & m_2\end{pmatrix} \begin{pmatrix} l_1 & l & l'_1 \\ -Q & 0 & Q\end{pmatrix} \begin{pmatrix} l_2 & l & l'_2 \\ -Q & 0 & Q\end{pmatrix}\label{eq:genmatrixelement}
\end{multline}	
\end{widetext}
where $\nu_l=1/\sqrt{Q}$.
The series can be summed numerically to obtain the Coulomb matrix elements. The series terminates at sufficiently large $l$ as the Wigner 3j symbols are nonzero only for those arguments that satisfy certain constraints (Sec 34.2 of Ref \onlinecite{NIST:DLMF}).

\subsection{Model interaction}
Matrix element (Eq. \ref{eq:defMatrixEl}) of the model interaction introduced in this work can be obtained by sandwiching the interaction shown in Eq.~5 of the main text between two-particle states. For convenience we reproduce the interaction in full detail here
\begin{multline}
\hat{V}=\sum_{l_1<l_2=0}^{\infty}\;\sum_{L=2Q+l_1+l_2-\delta_{l_1l_2}-2p+1}^{2Q+l_1+l_2-\delta_{l_1l_2}} \sum_{L_z=-L}^L V^{l_1,l_2}_L \times \\\times |l_1,l_2,L,L_z\rangle \langle l_1,l_2,L,L_z| 
\end{multline}
Here $|l_1,l_2,L,L_z\rangle$ is the unique two-particle state of total angular momentum quantum number $L$ and $z$-component angular momentum $L_z$, constructed from two particles in the $n_1=l_1-Q$ and $n_2=l_2-Q$ LLs. Note that we have restored the sum over $L_z$ which was omitted in Eq.~5 of the main text for brevity.
With this, the matrix element (Eq. \ref{eq:defMatrixEl}) between the two-particle states $| (l'_1,m'_1),(l'_2,m'_2)\rangle$ and $| (l_1,m_1),(l_2,m_2)\rangle$ is given by 
\begin{multline}
\delta_{(l_1,l_2)(l'_1,l'_2)}V_L^{l_1,l_2}C_{(l_1,m_1) (l_2,m_2)}^{L,L_z} C_{(l_1,m'_1) (l_2,m'_2)}^{L,L_z} 
\end{multline}
where $C_{(l_1,m_1) (l_2,m_2)}^{L,L_z}$ is the Clebsch Gordan coefficient.

\section{Spectra from Exact Diagonalization}
\label{SSecIV}

\begin{figure}
\includegraphics[width=\columnwidth]{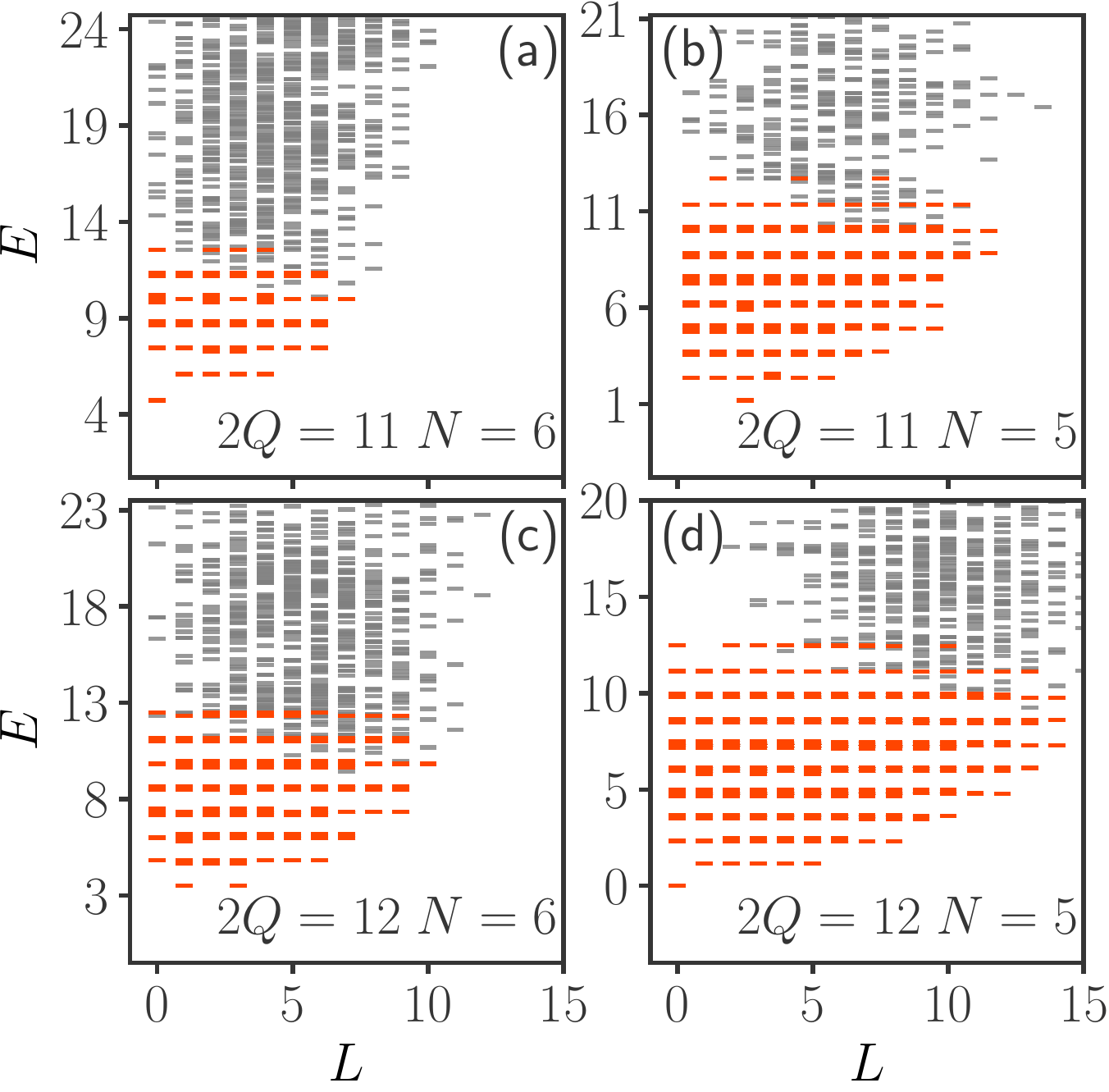}
\caption{Spectra for our model interaction for different $(N, 2Q)$ systems on the sphere. Orange markers indicate eigenstates with zero interaction energies. $\hbar\omega$ is set to $1$ and all non-zero pseudopotentials are set to 20. The black dashes show eigenstates that have nonzero interaction energy, i.e. are outside the ${\cal V}_\infty$ sector. \label{fig:moreModelSpec}}
\end{figure}

We have shown several spectra for our model Hamiltonian in the main text. 
Figure \ref{fig:moreModelSpec} shows additional representative spectra, all evaluated within the Hilbert space of the lowest 3 LLs. All nonzero pseudopotentials of the interaction are set to $20$ in units of $\hbar\omega_c=1$. The zero interaction energy states (i.e. the states belonging to $\mathcal{V}_\infty$) are indicated using orange markers and the remaining states are indicated with gray markers; the latter will be pushed to infinity in the limit where the interaction is taken to be infinitely strong. Panel (a) shows the system 
$(N,2Q)=(6,11)$, which maps into noninteracting fermion system $(N,2Q^*)=(6,1)$. Here, the spectrum contains a single incompressible ground state with $(N_0,N_1,N_2)=(2,4,0)$, followed by neutral excitations at $\hbar\omega_c$ at $L=1,2,3,4$ in the $(N_0,N_1,N_2)=(2,3,1)$ sector.  
Panel (b) shows the spectrum of $(N,2Q)=(5,11)$, which maps into $(N,2Q^*)=(5,3)$. The ground state corresponds to $(N_0,N_1,N_2)=(4,1,0)$, i.e. describes a single quasiparticle of the $1/3$ state. 
The angular momentum of this quasiparticle is predicted to be $L=5/2$. Panel (c) shows the spectrum of 
$(N,2Q)=(6,12)$, which maps into $(N,2Q^*)=(6,2)$. The ground state sector, with $(N_0,N_1,N_2)=(3,3,0)$,  has 
two quasiholes of the 2/5 state, each carrying an angular momentum of $l=2$. Addition of their angular momentum produces low energy states at total $L=1,3$, precisely as seen in panel (c). Panel (d) shows the $(N,2Q)=(5,12)$ which corresponds to the one filled LL state $(N,2Q^*)=(5,4)$. In all cases low energy states show same structure as what is expected in the LLL Coulomb spectrum.

\section{Counting of $\mathcal{V}_\infty$ states}
\label{SSecV}

In this section we present results comparing the number of numerically obtained zero energy ${\cal V}_\infty$ states $(N,2Q)$ with the dimension of the Hilbert space of the corresponding noninteracting system at $(N,2Q^*)$, with $2Q^*=2Q-2p(N-1)$. All exact diagonalizations are performed in the space of the lowest three LLs.

$\mathcal{V}_\infty$ states are constructed by acting $\hat{J}=\prod_{i<j=1}^N (\hat{Z}_i-\hat{Z}_j)^{2p}$ on the slater determinant states with a fixed number of particles $(N_0,N_1,N_2)$ in each LL. Maximum angular momenta of the particles in the slater determinant is less than that in the corresponding $\mathcal{V}_\infty$ state as the Jastrow factor $\hat{J}$ adds an angular momentum of $2p(N-1)$ for each particle. 

While the states constructed in this manner definitely belong to the $\mathcal{V}_\infty$ sector, one can ask: (i) Do different rearrangements of particles in a given $(N_0,N_1,N_2, \cdots)$ sector produce linearly independent $\mathcal{V}_\infty$ states? This seems likely, as the resulting states are very complex, but we do not have an analytic proof. (ii) Are there other other $\mathcal{V}_\infty$ states that are not captured by this construction? We have failed to find such states, but, rigorously speaking, cannot rule out this possibility a priori.  

We have conjectured that the states obtained in this fashion are all linearly independent and provide a complete basis for the $\mathcal{V}_\infty$ space. In other words, in the spherical geometry, the $\mathcal{V}_\infty$ eigenstates in exact diagonalization spectrum of $(N,2Q)$ have an exact one-to-one correspondence with the eigenstates of noninteracting fermions at $(N,2Q^*)$, with $2Q^* = 2Q - 2p(N-1)$. Alternatively: (i) all $\mathcal{V}_\infty$ states are of this form (or linear combinations of them), and (ii) $\mathcal{V}_\infty$ states constructed from different slater determinant states are linearly independent, i.e., the action of $\hat{J}$ does not not annihilate any wave function.

We provide convincing numerical evidence for this conjecture by explicitly diagonalizing our model interaction in various $(N_0,N_1,N_2)$ sectors at several values of $2Q$ to obtain their zero interaction energy eigenstates. The results are shown in the tables \ref{table:ZIE6}, \ref{table:ZIE5} and \ref{table:ZIE7}. In every case the number matches the number of different slater determinant states at $2Q^*$ in the $(N_0,N_1,N_2)$ sector. Only the number of the $L_z=0$ states is shown. In addition to the cases shown in the tables, the expected counting is obtained for the $(N,2Q)=(9,16)$ system, where only a single $\mathcal{V}_\infty$ state occurs in the sector $(N_0=1,N_1=3,N_2=5)$, corresponding to the $\nu=3/7$ ground state. The one-to-one correspondence is seen for all cases we have studied.

$L^2,L_z,N_0,N_1,N_2$ form good quantum numbers of the eigenstates of the model Hamiltonian. In the tables, we have have presented the comparison of the number of $\mathcal{V}_\infty$ states at each $(N_0,N_1,N_2)$ with the total number of non-interacting electronic eigenstates at $2Q^*$. $L_z$ was fixed at $0$ and $L^2$ was allowed to take any value.

In the Fig \ref{Sfig:178}, Fig \ref{Sfig:167} and Fig \ref{Sfig:166} we have shown a comparison of the noninteracting spectrum at $2Q^*$ and model Hamiltonian spectrum at $2Q$. The degeneracy of each orange dash($\mathcal{V}_\infty$ state) is shown next to it. The degeneracy indicates the number of independent states with a given energy and and given angular momentum $L$. In these plots $L_z$ is fixed at $0$ and $(N_0,N_1,N_2)$ can take any value. In all cases we find that the degenaracies in the spectra match identically.

Zero energy space dimensions were computed using Krylov-Schur algorithm\cite{stewart2002krylov} and setting large Krylov subspace dimensions.

\renewcommand{\arraystretch}{1.5}
\begin{center}
\begin{table}
\begin{tabular}{|c||c|c|}
\hline 
\multirow{1}{*}{$N_{0},N_{1},N_{2}$} & $2Q=11$ & $2Q=12$\tabularnewline
\hline 
0,0,5 & 6 & 12\tabularnewline
\hline 
0,1,4 & 41 & 74\tabularnewline
\hline 
0,2,3 & 80 & 146\tabularnewline
\hline 
0,3,2 & 56 & 108\tabularnewline
\hline 
0,4,1 & 14 & 31\tabularnewline
\hline 
0,5,0 & 1 & 3\tabularnewline
\hline 
1,0,4 & 29 & 56\tabularnewline
\hline 
1,1,3 & 130 & 246\tabularnewline
\hline 
1,2,2 & 165 & 322\tabularnewline
\hline 
1,3,1 & 70 & 147\tabularnewline
\hline 
1,4,0 & 9 & 21\tabularnewline
\hline 
2,0,3 & 35 & 76\tabularnewline
\hline 
2,1,2 & 103 & 222\tabularnewline
\hline 
2,2,1 & 79 & 178\tabularnewline
\hline 
2,3,0 & 17 & 40\tabularnewline
\hline 
3,0,2 & 13 & 36\tabularnewline
\hline 
3,1,1 & 23 & 64\tabularnewline
\hline 
3,2,0 & 9 & 26\tabularnewline
\hline 
4,0,1 & 1 & 5\tabularnewline
\hline 
4,1,0 & 1 & 5\tabularnewline
\hline 
5,0,0 & 0 & 1\tabularnewline
\hline 
\end{tabular}
\caption{The number of independent $L_z=1/2$ zero interaction energy ($\mathcal{V}_\infty$) eigenstates in exact-diagonalization spectra for many different $(N_0,N_1,N_2)$ sectors for several values of $2Q$.  The total number of particles is $N=5$. In every single case, the number of numerically obtained zero interaction energy eigenstates matches exactly with the number of slater determinant states for noninteracting electrons in the $(N_0,N_1,N_2)$ sector at $2Q^*=2Q-2(N-1)$.
\label{table:ZIE5}
}
\end{table}
\end{center}

\renewcommand{\arraystretch}{1.5}
\begin{center}
\begin{table}
\begin{tabular}{|p{2 cm}||c|c|c|c|}
\hline
{$N_{0},N_{1},N_{2}$} & $2Q=11$ & $2Q=12$ & $2Q=15$ & $2Q=16$\tabularnewline
\hline 
6,0,0 & 0 & 0 & 1 & 1\tabularnewline
\hline 
5,1,0 & 0 & 0 & 6 & 19\tabularnewline
\hline 
4,2,0 & 0 & 0 & 44 & 108\tabularnewline
\hline 
5,0,1 & 0 & 0 & 6 & 21\tabularnewline
\hline 
3,3,0 & 0 & 2 & 106 & 236\tabularnewline
\hline 
4,1,1 & 0 & 0 & 106 & 265\tabularnewline
\hline 
2,4,0 & 1 & 3 & 100 & 210\tabularnewline
\hline 
3,2,1 & 0 & 10 & 452 & 978\tabularnewline
\hline 
4,0,2 & 0 & 0 & 59 & 143\tabularnewline
\hline 
1,5,0 & 0 & 1 & 34 & 74\tabularnewline
\hline 
2,3,1 & 4 & 28 & 656 & 1328\tabularnewline
\hline 
3,1,2 & 0 & 13 & 550 & 1165\tabularnewline
\hline 
0,6,0 & 0 & 0 & 4 & 8\tabularnewline
\hline 
1,4,1 & 2 & 15 & 334 & 674\tabularnewline
\hline 
2,2,2 & 14 & 74 & 1372 & 2662\tabularnewline
\hline 
3,0,3 & 0 & 5 & 186 & 393\tabularnewline
\hline 
0,5,1 & 0 & 1 & 48 & 102\tabularnewline
\hline 
1,3,2 & 18 & 74 & 1086 & 2054\tabularnewline
\hline 
2,1,3 & 12 & 61 & 1028 & 1971\tabularnewline
\hline 
0,4,2 & 3 & 13 & 236 & 452\tabularnewline
\hline 
1,2,3 & 34 & 118 & 1396 & 2556\tabularnewline
\hline 
2,0,4 & 3 & 13 & 236 & 452\tabularnewline
\hline 
0,3,3 & 12 & 40 & 470 & 858\tabularnewline
\hline 
1,1,4 & 18 & 61 & 708 & 1292\tabularnewline
\hline 
0,2,4 & 14 & 40 & 410 & 728\tabularnewline
\hline 
1,0,5 & 2 & 9 & 114 & 214\tabularnewline
\hline 
0,1,5 & 4 & 13 & 146 & 264\tabularnewline
\hline 
0,0,6 & 1 & 1 & 18 & 32\tabularnewline
\hline 
\end{tabular}
\caption{The number of independent $L_z=0$ zero interaction energy ($\mathcal{V}_\infty$) eigenstates in exact-diagonalization spectra for many different $(N_0,N_1,N_2)$ sectors for several values of $2Q$.  The total number of particles is $N=6$. In every single case, the number of numerically obtained zero interaction energy eigenstates matches exactly with the number of slater determinant states for noninteracting electrons in the $(N_0,N_1,N_2)$ sector at $2Q^*=2Q-2(N-1)$.
\label{table:ZIE6}}
\end{table}
\end{center}

\renewcommand{\arraystretch}{1.5}
\begin{center}
\begin{table}
\begin{tabular}{|c|c||c|c|}
\hline 
$N_{0},N_{1},N_{2}$ & $2Q=16$ & $N_{0},N_{1},N_{2}$ & $2Q=16$\tabularnewline
\hline 
\hline 
7,0,0 & 0 & 0,2,5 & 210\tabularnewline
\hline 
0,7,0 & 1 & 3,0,4 & 108\tabularnewline
\hline 
0,0,7 & 4 & 1,5,1 & 91\tabularnewline
\hline 
6,1,0 & 0 & 0,4,3 & 236\tabularnewline
\hline 
6,0,1 & 0 & 0,3,4 & 344\tabularnewline
\hline 
1,6,0 & 5 & 1,1,5 & 356\tabularnewline
\hline 
0,6,1 & 7 & 4,2,1 & 91\tabularnewline
\hline 
1,0,6 & 38 & 4,1,2 & 114\tabularnewline
\hline 
0,1,6 & 52 & 2,4,1 & 286\tabularnewline
\hline 
5,2,0 & 3 & 1,4,2 & 520\tabularnewline
\hline 
5,0,2 & 4 & 3,3,1 & 286\tabularnewline
\hline 
2,5,0 & 26 & 2,1,4 & 700\tabularnewline
\hline 
0,5,2 & 68 & 1,2,4 & 1016\tabularnewline
\hline 
4,3,0 & 21 & 3,1,3 & 482\tabularnewline
\hline 
5,1,1 & 7 & 1,3,3 & 1136\tabularnewline
\hline 
2,0,5 & 108 & 3,2,2 & 628\tabularnewline
\hline 
3,4,0 & 40 & 2,3,2 & 1022\tabularnewline
\hline 
4,0,3 & 38 & 2,2,3 & 1372\tabularnewline
\hline 
\end{tabular}
\caption{The number of independent $L_z=0$ zero interaction energy ($\mathcal{V}_\infty$) eigenstates in exact-diagonalization spectra for different $(N_0,N_1,N_2)$ sectors within the $(N,2Q)=(7,16)$ system. In every single case, the number of numerically obtained zero interaction energy eigenstates matches exactly with the number of slater determinant states of noninteracting electrons in the $(N_0,N_1,N_2)$ sector at $2Q^*=2Q-2(N-1)$.
\label{table:ZIE7}}
\end{table}

\begin{figure*}
\includegraphics[width=\textwidth]{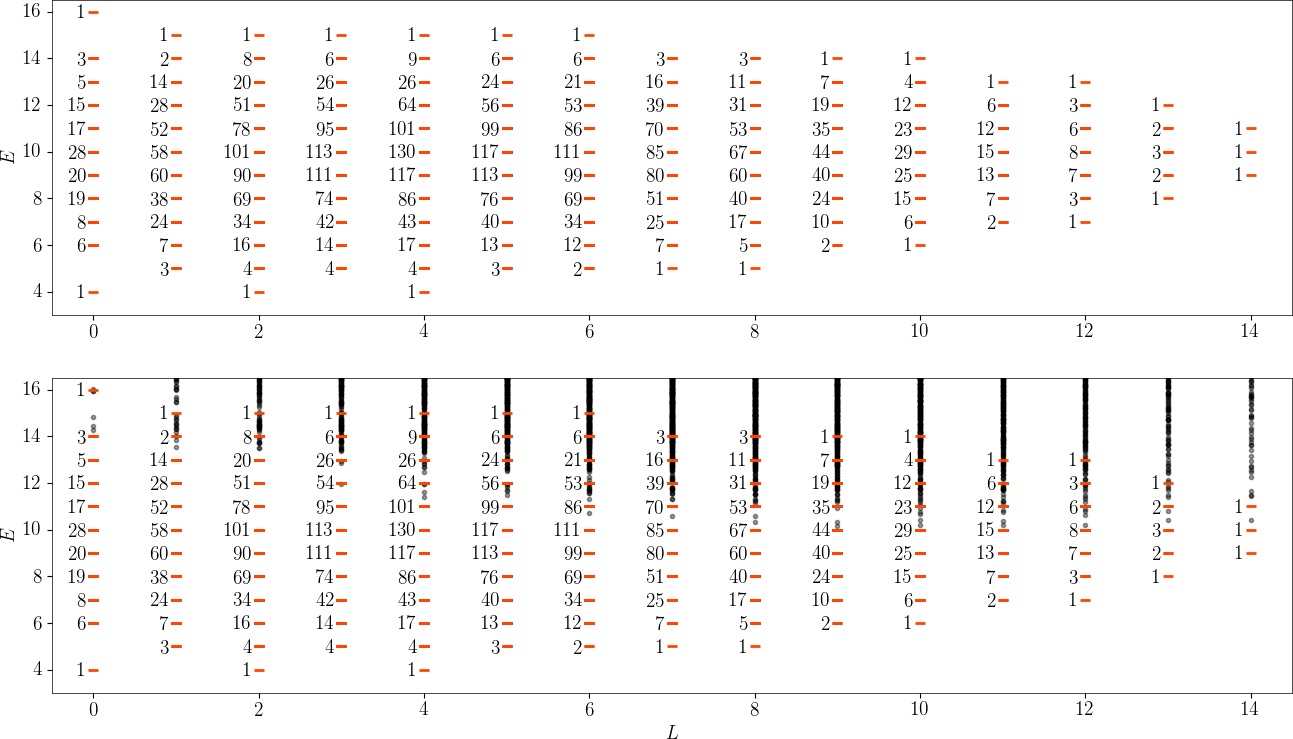}
\caption{(top): Spectrum of $N=8$ non-interacting electrons at flux $2Q^*=3$ in the Hilbert space of three LLs, with the cyclotron energy set to $\hbar\omega=1$. (bottom) Spectrum of the model Hamiltonian for $N=8$ electrons at flux $2Q=17$ in the Hilbert space of three LLs, with the cyclotron energy set to $\hbar\omega=1$. All interaction pseudopotentials are set to 20. Orange dashes show $\mathcal{V}_\infty$ states and the black dots shown the finite interaction energy states. The numbers next to the orange dashes show the degeneracy of each line. The flux values are related by $2Q^*=2Q-2(N-1)$. In every case the degeneracies in the two spectra match. LL energies have been taken to be $n\hbar\omega$ for simplicity. \label{Sfig:178} }
\end{figure*}

\begin{figure*}
\includegraphics[width=\textwidth]{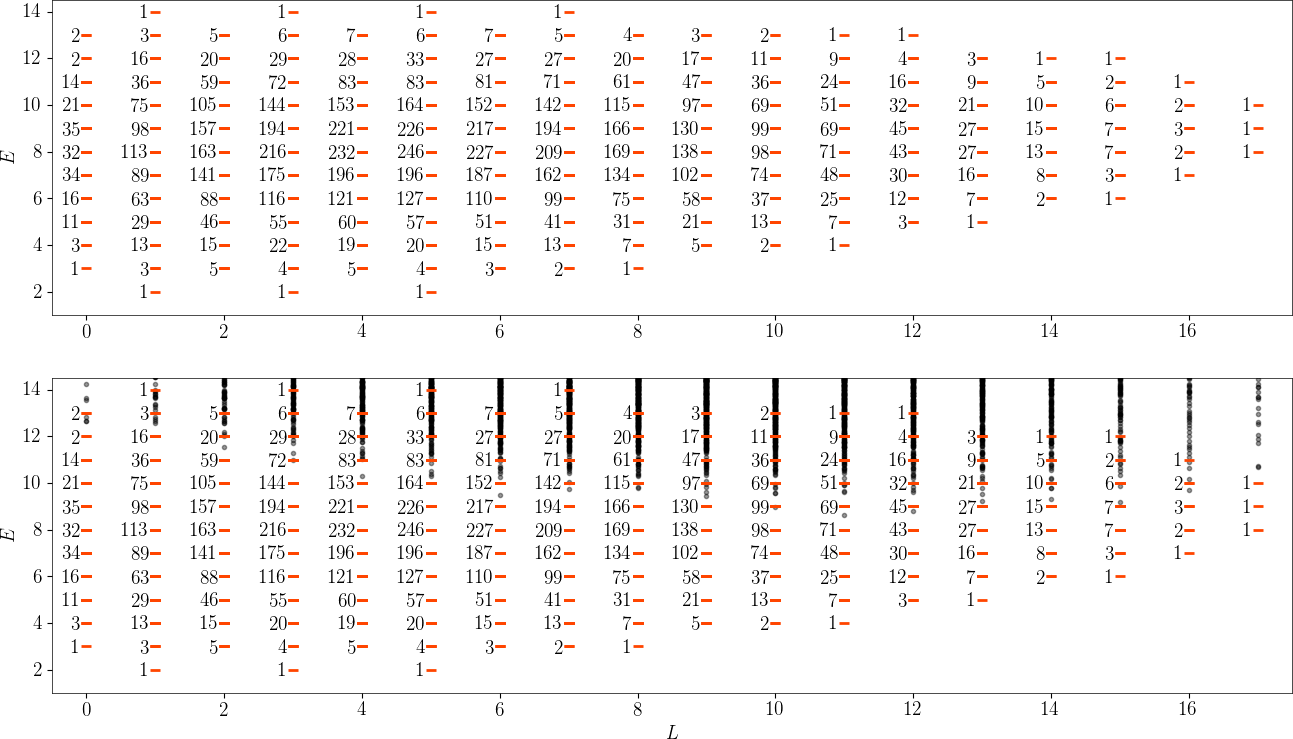}
\caption{Similar to Fig.~\ref{Sfig:178} but for a system of size $N=7$. Bottom panel shows the spectrum of the model interaction at $2Q=16$ and the top panel shows the spectrum of non-interacting electrons at $2Q^*=2Q-2(N-1)=4$.  \label{Sfig:167} }
\end{figure*}

\begin{figure*}
\includegraphics[width=\textwidth]{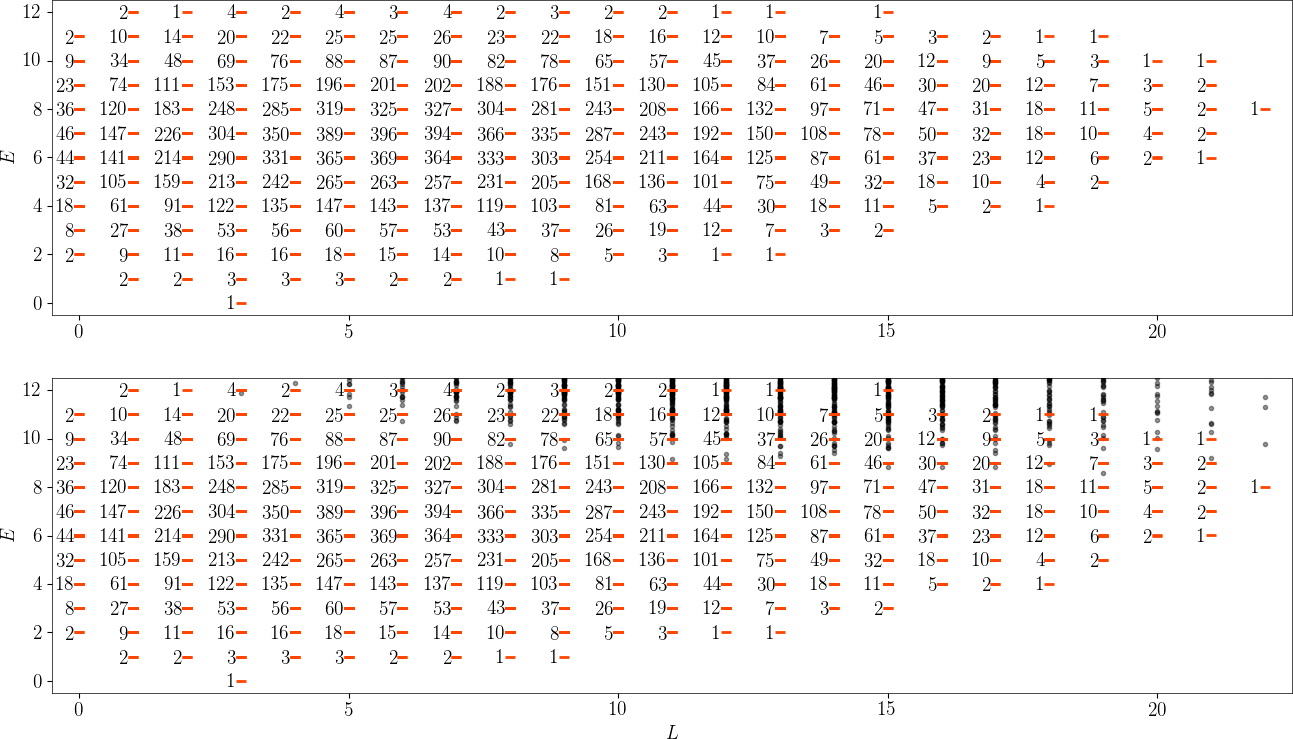}
\caption{Similar to Fig.~\ref{Sfig:178} but for a system of size $N=6$. Bottom panel shows the spectrum of the model interaction at $2Q=16$ and the top panel shows the spectrum of non-interacting electrons at $2Q^*=2Q-2(N-1)=6$.\label{Sfig:166}}
\end{figure*}

\end{center}

\section{Adiabatic continuity at $\nu=1/3$}
\label{SSecVI}

\begin{figure}
\includegraphics[width=\columnwidth]{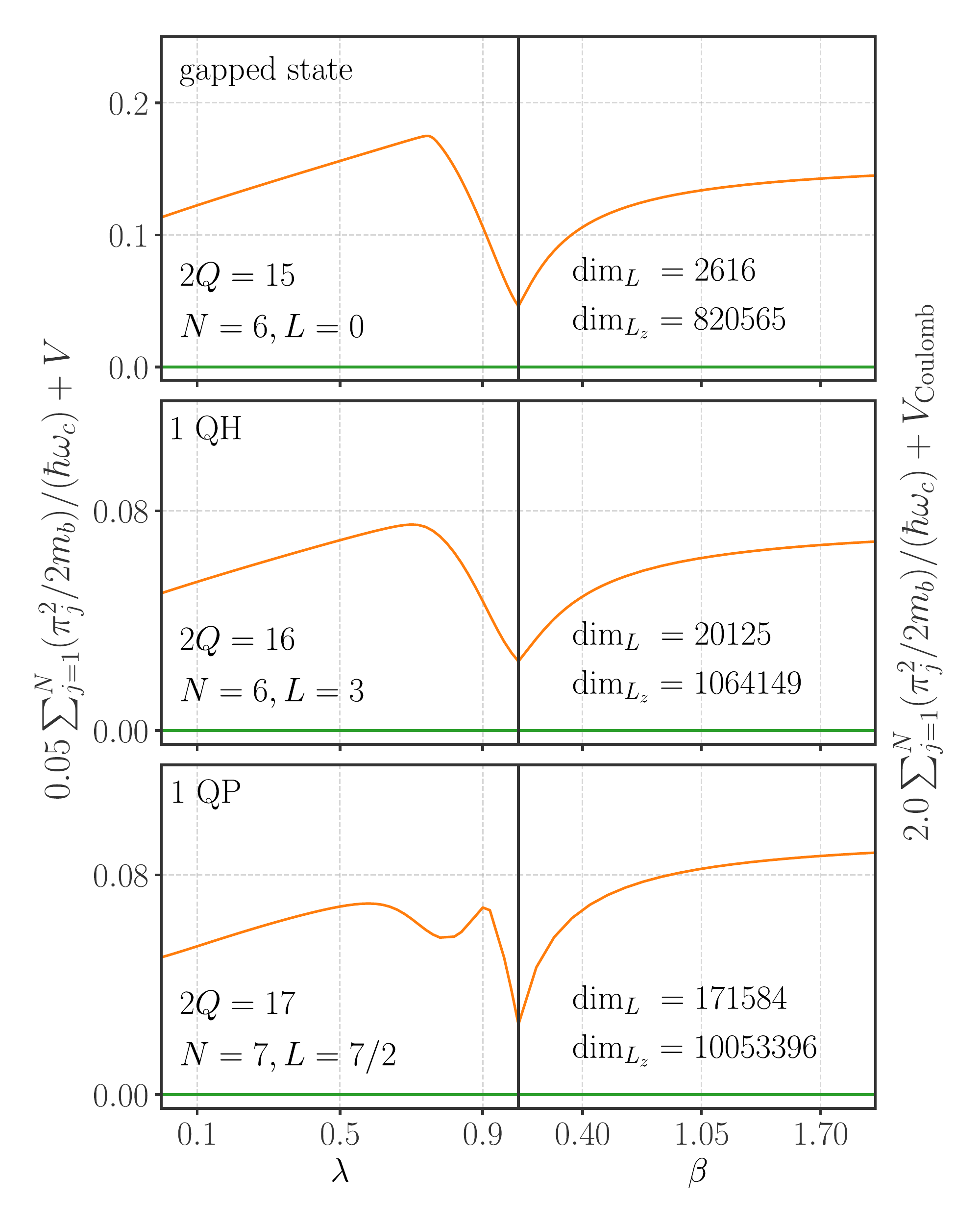}
\caption{Demonstration of adiabatic connectivity of GS (top), single QH (middle) and single QP (bottom) states of our model Hamiltonian and the those of the LLL Coulomb Hamiltonian for a finite system near
$\nu=1/3$. The form of Hamiltonian is $\hat{H}=\beta \sum^{N}_{j=1}(\hat{\boldsymbol{\pi}}^{2}_{j}/2m_{b})/(\hbar \omega_{c}) + (1-\lambda)\hat{V} + \lambda \hat{V}_{\rm Coulomb}$. Three LLs were included in the calculation. The left panels show the change in the spectra as $\lambda$ changes from $0$ to $1$ keeping $\beta=0.05$ and the right panels show the variation as $\beta$ changes from $0.05$ to $2$. Note that the single QP and single QH states have an angular momentum of $L=N/2+1$ and $N/2$ respectively.
All the energies are relative to the GS in the corresponding system.\label{fig:adiabatic13rd}}
\end{figure}

In the main text, we have shown how the ground state, single QP and single QH states of our model Hamiltonian at $\nu=2/5$ evolves into the corresponding states of the Coulomb interaction in the LLL under a continuous deformation of the Hamiltonian. We demonstrate the same for $\nu=1/3$ in this section.

Figure \ref{fig:adiabatic13rd} shows how the lowest two states of the spectrum of the model Hamiltonian (in the spherical geometry) evolves as the Hamiltonian is continuously changed to the LLL Coulomb Hamiltonian along a particular path in the parameter space. Each row of the figure shows the evolution of the spectrum within a single total angular momentum sector. The Hamiltonian is parametrized by two parameters $\lambda$ and $\beta$ as follows
\begin{equation}
\hat{H}(\beta,\lambda) = \frac{\beta}{\hbar \omega_c} \sum_{i=1}^N \frac{\hat{\boldsymbol \pi}_i^2}{2m} + (1-\lambda)\hat{V} + \lambda \hat{V}_{\rm Coulomb}
\end{equation}
where $\hat{V}$ is the model Hamiltonian and $\hat{V}_{\rm Coulomb}$ is the Coulomb Hamiltonian. All energies are quoted in units of $e^2/(\epsilon \ell)$. Nonzero pseudopotentials of the model Hamiltonian are all set to $1$.

\begin{figure}
\includegraphics[width=\columnwidth]{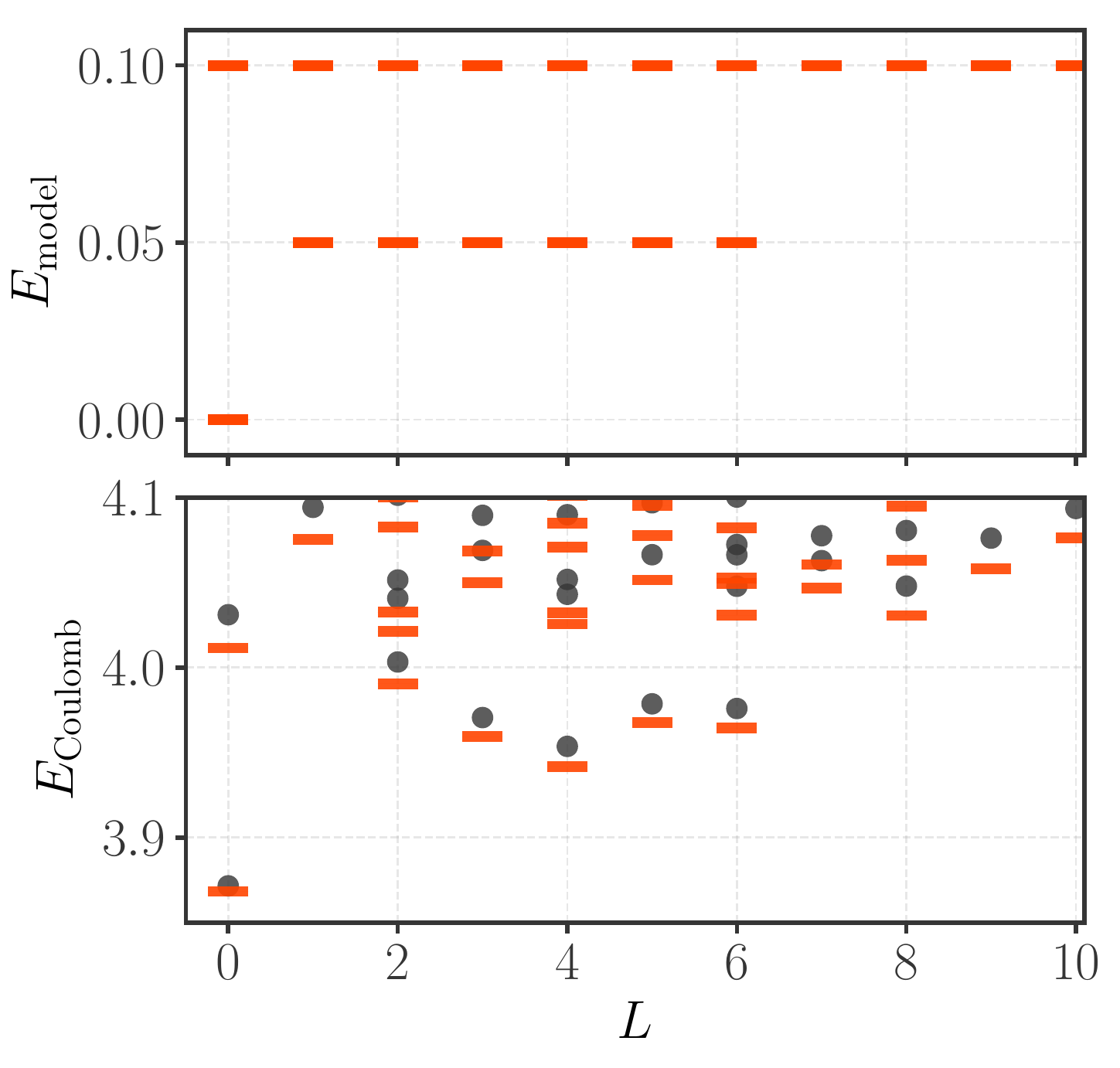}
\caption{Low energy spectra of the Hamiltonians $\hat{H}(\beta=0.05,\lambda=0)$ 
(top) and $\hat{H}(\beta=2.0,\lambda=1)$ (bottom,dashes) for a systen of $N=6$ particles and $2Q=15$. Dots in the bottom figure represent the LLL Coulomb spectrum.
For simplicity the subleading $1/Q$ dependent part of the cylotron energy on the sphere was omitted in the top panel. 
\label{fig:spectra13rd}}
\end{figure}

The left panels show the change in the spectrum as $\lambda$ is changed from $0$ to $1$ with $\beta=0.05$. At the left end, the Hamiltonian $\hat{H}(\beta =0.05,\lambda= 0)$ represents a system with a small cyclotron gap of $0.05$ and particles interacting via the model Hamiltonian $\hat{V}$. At the right end of the left panels, $\hat{H}(\beta = 0.05,\lambda = 1)$ represents the Hamiltonian of the system with the same cyclotron gap but instead interacting via the Coulomb interaction.  The right panels show the spectral transformation as $\beta$ is varied from $0.05$ to $2$ keeping $\lambda=1$. 
At the rightmost end, on account of the relatively large cyclotron energy, particles predominantly reside in the LLL. The spectrum here closely match the spectrum of the Coulomb interaction for particles strictly confined to the LLL, which corresponds to the case of infinite cyclotron gaps (Fig \ref{fig:spectra13rd} (bottom)). 
The uniform ground state of the model Hamiltonian ($L=0$ sector shown in the top) at $2Q=3N-3$ is identical to the Laughlin state, and we expect that it is adiabatically connected to the Coulomb Hamiltonian. We do find that along the path that we have chosen, the two are indeed connected without gap closing (Fig \ref{fig:adiabatic13rd} top).As discussed below, the simplest low energy states namely single QH and QP states also appear to be adiabatically connected to each other. 
(Fig \ref{fig:spectra13rd})

Single quasihole state occurs with an angular momentum of $L=N/2$ in a system with $2Q=3N-2$. This state is again identical to the single QH state of the Laughlin state and we do find that along the path that we have studied the single QH of the model and Coulomb interactions are connected without gap closing (Fig \ref{fig:adiabatic13rd} middle).

Most importantly, the single QP of the model which occurs with angular momentum $L=N/2+1$ in a system $2Q=3N-4$ is also adiabatically connected to the corresponding state of the Coulomb Hamiltonian (Fig \ref{fig:adiabatic13rd} bottom)

\section{Topological properties}
\label{SSecQP}

Many topological properties follow from the counting of the low energy levels, which we know exactly for our model through its one to one correspondence with the noninteracting problem of electrons at an effective filling factor. We discuss these in some detail in this section.

\underline{Fractional charge} As reviewed in Ref.~\cite{Jain07}, the fractional charge for the quasiparticle (quasihole) can be obtained from an analogy to the noninteracting system. Essentially, one asks the question: How many quasiparticles (quasiholes) are created when one adds a single electron to (removes a single electron from) the system? Following verbatim the discussion in Ref.~\cite{Jain07}, $2pn+1$ quasiparticles are created when one electron is added to the system, thus producing a charge of magnitude $e^*=1/(2pn+1)$, in units of the electron charge, for each quasiparticle. 


The charge can also be obtained from the density profile. We consider here the single quasiparticle state of $2/5$ for $N=9$, which occurs for a system with $2Q=18$ at an angular momentum of $L=3$. When placed in an $L_z=L$ state, the excess charge is maximally moved to one of the poles of the system. The net charge accumulated due to the quasiparticle can be calculated as the saturation value (at large $\theta$) of
\begin{equation}
Q(\theta) = \int_0^\theta \left [\rho(x) - {\rho}(\pi) \right] 2\pi\sin x dx
\end{equation}
where $\theta$ is the polar angle, $\rho(x)$ is the local charge density at a polar angle $x$ (which will be azimuthally symmetric, ie independent of azimuthal angle $\phi$) and ${\rho}(\pi)$ is the density at the south pole.

\begin{figure}
\includegraphics[width=\columnwidth]{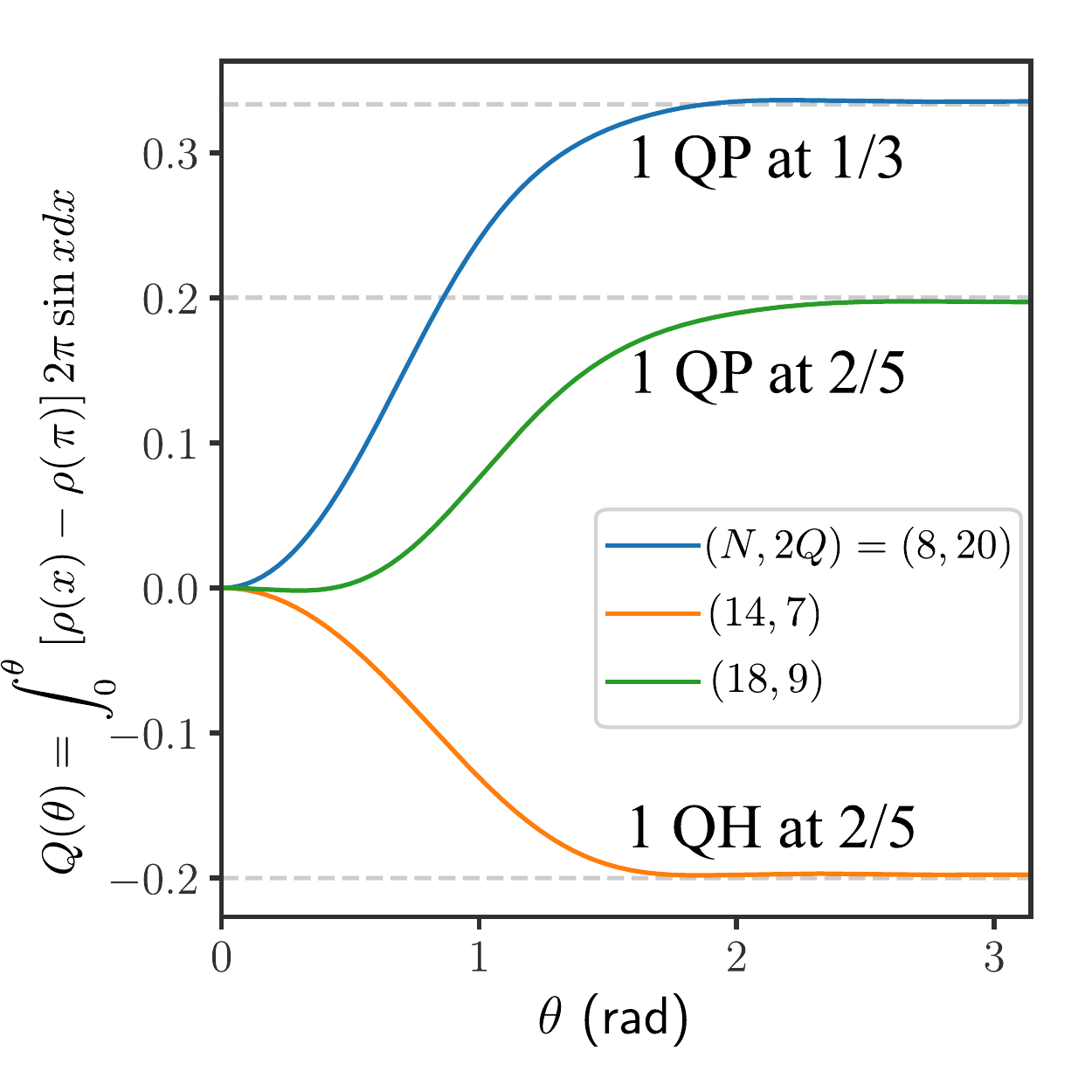}
\caption{
Cumulative charge inside a region defined by angle $\theta$ for a QP and a QH of the 2/5 FQH state and for a QP of the 1/3 FQH state. The plot shows excess charge near the north pole (measured relative to the density at the south pole) in a system where there is a single QP/QH placed in a highest weight ($L_z=L$) state, corresponding to the QP/QH located at the north pole. In all cases we find that the magnitude of the net charge in the around the north pole is $1/(2p\nu^*+1)$ in units of the electron charge.
\label{SFig:charge}}
\end{figure}

Fig \ref{SFig:charge} shows the result for $Q(\theta)$ indicating that the total charge accumulated around the pole is $1/5$ as expected. Similar analysis of the QH of $2/5$ shows its charge to be $1/5$ in magnitude again. 

The single QH state of $1/3$ in our model is identical to that of the Laughlin quasihole state and therefore has a charge of $1/3$. The single quasiparticle is found to have a charge of $1/3$ as shown by the blue line of the figure, evaluated for a system of 8 electrons at $2Q=20$.

\underline{Fractional statistics}: The one-to-one analogy between our solutions and the integer quantum Hall problem for states consisting of several QPs or QHs indicates that these obey Abelian statistics. This follows because, just as for the integer quantum Hall state, specifying the positions produces a unique wave function, and thus any braidings may only produce a phase factor. Furthermore, general arguments discussed in Ref.~\onlinecite{Jain07} also fix the value of the statistics.

\underline{Edge states}: The counting of edge excitations of the $n/(2pn+1)$ states in our model is identical to that of the $\nu^*=n$ integer quantum Hall state. This implies that the edge excitations for our model system are exactly described in terms of $n$ chiral bosonic modes, which is also believed to be the case for the lowest Landau level FQH states at these filling factors. 

\section{Trugman-Kivelson Hamiltonian}
\label{SSecVIII}

Trugman and Kivelson~\cite{Trugman85} (TK) introduced the interaction 
\begin{equation}
V_{\rm TK}(r)=\nabla^2 \delta(r),
\end{equation} 
which obtains Laughlin's wave function as the unique solution at $\nu=1/3$ in the LLL Hilbert space. Interestingly, if one 
sets the lowest two LLs degenerate while sending the remaining higher LLs to infinity, then the unprojected Jain state at $\nu=2/5$:
\begin{equation}
\Psi^{\rm CF}_{2/5} = \prod_{i<j=1}^{N} (z_i-z_j)^{2} \;\Phi_{2}(\{z_i\}),
\label{eq:CF25}
\end{equation}
where $\Phi_{2}(\{z_i\})$ is the spin polarized integer quantum Hall state at filling fraction $2$, appears as the unique zero-energy ground state for the TK interaction~\cite{Jain90b,Jain90,Rezayi91}. This state occurs at flux $2Q=5N/2-4$ on the sphere. It is interesting to ask in what sense the TK interaction is different from the one considered here.

On the sphere, the TK interaction can be expanded in spherical Harmonics as (compare with Eq. \ref{eq:multipoleexp})
\begin{equation}
\nabla^2 \delta({{\bf r}_a-{\bf r}_b}) = -\sum_{l=0}^\infty\sum_{m=-l}^l \frac{l(l+1)}{Q} Y_{0lm} (a) Y^*_{0lm} (b)
\end{equation}
where we have expanded the Dirac delta function in monopole spherical harmonics (Eqn 1.17.25 of Ref \onlinecite{NIST:DLMF}). The matrix elements of the TK Hamiltonian can be found in a manner similar to that for the Coulomb Hamiltonian and have the same form as Eq.~\ref{eq:antisymmet} and Eq. \ref{eq:genmatrixelement} but with $\nu_l = -{l(l+1)(2l+1)}/{4\pi Q}$.

It is interesting to ask in what way the TK Hamiltonian is different from ours. To address this, we consider the system of particles living in the Hilbert space of the two lowest LLs. 
Our model Hamiltonian introduced in this work imposes an energy penalty on (i) all two-particle states of relative angular momenta $L=2Q+1$ and $2Q$ on the sphere; and on (ii) the unique multiplet of two-particle states of angular momenta $2Q-1$ in the LLL.

Like our model Hamiltonian, the TK Hamiltonian also imposes an energy penalty on all states in the $L=2Q+1$ and $2Q$ sectors. In the $L=2Q-1$ sector, however, it imposes a penalty on 
a particular two-particle multiplet in the three dimensional (multiplet) space of the $L=2Q-1$ states. This multiplet is a linear combination of two-particles states in multiple LLs. That makes TK Hamiltonian LL non-conserving. The ``forbidden" two-particle multiplet of the TK Hamiltonian in the $L=2Q-1$ sector can be inferred from numerically studying the two-particle spectrum of the TK Hamiltonian, using the matrix elements evaluated as described above.

\begin{figure}
\vspace{10pt}
\includegraphics[width=\columnwidth]{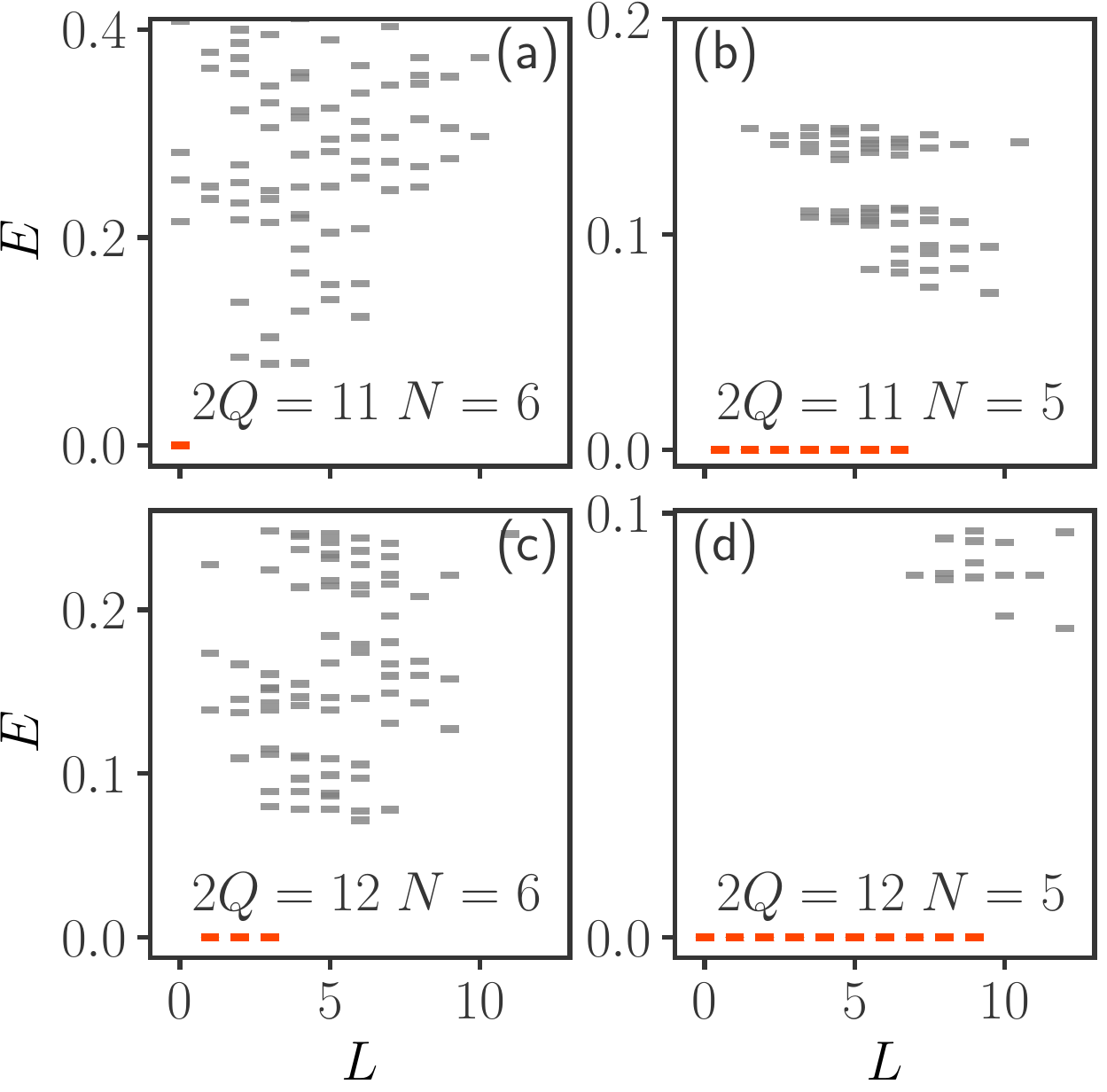}
\caption{Spectra for the Trugman-Kivelson interaction for several $(N,2Q)$ systems (same as in Fig.~\ref{fig:moreModelSpec}) obtained by exact diagonalization. Only two lowest Landau levels are included, which are assumed to be degenerate. Zero energy states are shown in orange. The energy is quoted in arbitrary units (set by the KT interaction), but all four plots are in the same units.\label{fig:moreTKSpec}}
\end{figure}

Figure~\ref{fig:moreTKSpec} depicts spectra for the TK Hamiltonian for the same $(N,2Q)$ systems as in Fig.~\ref{fig:moreModelSpec}. Only the two lowest LLs (taken as degenerate) are included in the diagonalization. The unique zero energy state occurring in the panel (a) is identical to the Jain unprojected wave function in Eq.~\ref{eq:CF25}. The zero energy states in the remaining panels are highly degenerate; their counting matches with that of the corresponding $(N,2Q^*)$ system of noninteracting fermions with the lowest two LLs taken as degenerate.
\vspace{25pt}
\bibliography{supplement.bib}
\bibliographystyle{apsrev}

\end{document}